\documentclass[conference]{IEEEtran}
\usepackage{graphicx}
\usepackage{tabularx,booktabs}
\usepackage{dingbat}
\usepackage{diagbox}
\usepackage{multirow} 
\usepackage[hyphens]{url}
\usepackage[hidelinks,breaklinks]{hyperref}
\usepackage{xcolor}
\usepackage{amsfonts, amsmath, amsthm, amssymb}
\usepackage{subcaption}
\hypersetup{breaklinks=true}
\usepackage{tikz}
\usepackage{textcomp}
\usepackage{lipsum}
\usepackage{booktabs}
\usepackage{multirow}
\usepackage{setspace}
\IEEEoverridecommandlockouts
\newcommand\copyrighttext{%
  \footnotesize \textcopyright 2024 IEEE.  Personal use of this material is permitted.  Permission from IEEE must be obtained for all other uses, in any current or future media, including reprinting/republishing this material for advertising or promotional purposes, creating new collective works, for resale or redistribution to servers or lists, or reuse of any copyrighted component of this work in other works.}
\newcommand\copyrightnotice{%
\begin{tikzpicture}[remember picture,overlay]
\node[anchor=south,yshift=10pt] at (current page.south) {\fbox{\parbox{\dimexpr\textwidth-\fboxsep-\fboxrule\relax}{\copyrighttext}}};
\end{tikzpicture}%
}
\title{Real-World Performance Evaluations of Low-Band 5G NR/4G LTE 4×4 MIMO on Commercial Smartphones}
\author{
    \IEEEauthorblockN{Pasapong Wongprasert\IEEEauthorrefmark{1}, Kasidis Arunruangsirilert\IEEEauthorrefmark{2}, Jiro Katto\IEEEauthorrefmark{2}} 
    \IEEEauthorblockA{\IEEEauthorrefmark{1}Department of Electrical Engineering, Chulalongkorn University, Bangkok, Thailand}
    \IEEEauthorblockA{\IEEEauthorrefmark{2}Department of Computer Science and Communications Engineering, Waseda University, Tokyo, Japan
    \\6670331021@student.chula.ac.th, \{kasidis, katto\}@katto.comm.waseda.ac.jp}
}

\begin{document}

\maketitle
\copyrightnotice
\begin{abstract} 
All 3GPP-compliant commercial 5G New Radio (NR)-capable UEs on the market are equipped with 4×4 MIMO support for Mid-Band frequencies (\textgreater1.7 GHz) and above, enabling up to rank 4 MIMO transmission. This doubles the theoretical throughput compared to rank 2 MIMO and also improves reception performance. However, 4×4 MIMO support on low-band frequencies (\textless1 GHz) is absent in every commercial UEs, with the exception of the Xperia 1 flagship smartphones manufactured by Sony Mobile and the Xiaomi 14 Pro as of January 2024. The reason most manufacturers omit 4×4 MIMO support for low-band frequencies is likely due to design challenges or relatively small performance gains in real-world usage due to the lack of 4T4R deployment on low-band by mobile network operators around the world.

In Thailand, 4T4R deployment on the b28/n28 (APT) band is common on True-H and dtac networks, enabling 4×4 MIMO transmission on supported UEs. In this paper, the real-world 4×4 MIMO performance on the b28/n28 (APT) band will be investigated by evaluating the reliability test under different signal conditions and the maximum throughput test by evaluating the performance under optimal conditions, using the Sony Xperia 1 III and the Sony Xperia 1 IV smartphone. Devices from other manufacturers are also used in the experiment to investigate the performance with 2Rx antennas for comparison. Through firmware modifications, the Sony Xperia 1 III and IV can be configured to use only 2 Rx ports on low-band, enabling the collection of comparative 2 Rx performance data as a reference.

\end{abstract}

\begin{IEEEkeywords}
5G New Radio (NR), 4G Long Term Evolution (LTE), MIMO, User Equipment (UE), Radio Access Network, Wireless Communication, low-band.
\end{IEEEkeywords}
\vspace{-1mm}
\section{Introduction}
\setstretch{0.95}
Prior to the commercialization of 5G NR (New Radio) capable User Equipment (UE), 4×4 MIMO support was primarily featured in high-end flagship devices only [2]. However, with the advent of the 5G era, support for 4×4 MIMO is now implemented in every 3GPP-compliant commercial NR-capable UEs, including entry-level devices with rare exceptions, such as the iPhone SE (2022) \cite{AppleiPhoneSESpecs}. The reason for this widespread adoption can be explained due to the fact that 4×4 MIMO support is made mandatory for a set of NR bands, which also happen to be the most commonly allocated bands for 5G around the world. According to the technical standard TS 138 306 from ETSI based on the 3GPP recommendation TS38.306 Release 15, which defines the capabilities of the 5G NR UEs, the document states that UEs are required to be equipped with a minimum of 4 Rx antenna ports for n7, n38, n41, n77, n78 and n79 NR bands \cite{3GPP_36-306}. This mandated compliance has also resulted in a consequential benefit observed in closely related frequency bands such as n1, n3, and n40 as the antennas assigned for the mandated 4Rx bands can be utilized for these bands as well due to their close frequency ranges. Consequently, nearly all NR-capable UEs, including entry-level devices, include 4×4 MIMO support for these additional bands, despite not being explicitly mandated.

Another frequency band that is commonly allocated for 5G NR is n28 (700 MHz). Support for 4×4 MIMO on low-band frequencies such as band 8, band 20, and band 28 is absent in virtually every commercial NR UEs except for the Xperia 1 flagship smartphones manufactured by Sony and the recently released Xiaomi 14 Pro smartphone, as of January 2024. This general lack of support is likely attributed to design challenges, as the antennas designed for low-band require larger dimensions to achieve sufficient efficiency, consequently occupying more internal space. This may lead to the necessity for design compromises \cite{Bancroft2004FundamentalDL}.

The utilization of 4 Rx antennas in UEs can enhance the user experience, in a way similar to the utilization of 2 Tx antennas found in some of the high-end 5G UEs \cite{10118777}. Firstly, it increases the maximum number of MIMO layers and in this case, the configuration enables up to 4 layers of MIMO transmission (Rank 4 MIMO), effectively doubling the theoretical throughput compared to 2-layer MIMO. Moreover, the additional antennas can enhance receive diversity gain by increasing redundancy and thereby improving reliability \cite{huawei20204x4mimo}. In both 4G LTE and 5G NR radio access technologies (RAT), MIMO rank is negotiated by the UE and base station through a sequence of signaling exchange between the evolved Node B (eNB)/next Generation Node B (gNB) and the UE. First, the UE reports the Channel State Information (CSI) which contains the Channel quality index (CQI), Precoding Matrix Indicator (PMI), and the preferred Rank Indicator (RI) to the eNB/gNB. The eNB/gNB then decides the transmission mode and MIMO rank based on the CSI reported by the UE along with factors that are specific to the eNB/gNB such as scheduling constraints and load. The eNB/gNB then pushes the Downlink Control Information (DCI) message to inform the UE of the final RI assignment. Next, the modulation coding scheme (MCS) negotiation process also follows the same process however, If the Block Error Rate (BLER) is lower than anticipated, a higher MCS will be negotiated until the 10 percent BLER target can no longer be maintained \cite{3GPP_36-212, 3GPP_38-212}. In both 4G LTE and 5G NR systems, MIMO transmission can be categorized into three scenarios. When the number of transmission ports at the eNB/gNB matches the number of utilized Rx ports of the UE, and the rank is equal to the maximum number of available antennas, the system operates in pure spatial multiplexing. In this mode, throughput is maximized and is suitable when channel quality is adequate. When the rank or MIMO layer is equal to one and more than one antenna is utilized either on the transmit (Tx) or receive (Rx) side, the system operates in pure transmit diversity or pure receive diversity respectively. This mode aims to maximize link reliability when the channel quality is poor. If the RI is greater than one but less than the maximum number of antennas in either the Tx or Rx side, then there is a concurrent utilization of both spatial multiplexing and transmit diversity, allowing a balance of throughput and reliability \cite{3GPP_36-213}. \looseness=-1

In this research, the real-world performance of low-band 4×4 MIMO will be investigated in different scenarios with the Sony Xperia 1 smartphones. To provide a more comprehensive evaluation, other commercial smartphones without 4×4 MIMO support on low-band will also be tested for comparison in order to evaluate and understand the impact of having 2 additional Rx antennas.

\section{Experimental procedure}

\subsection{Network environment}

The entire experiment will be carried out in a home environment where the UEs will be placed on a stand to ensure that all UEs have a consistent orientation and positioning throughout the experiment as shown in Figure~\ref{fig:testSetup}. There are a total of 16 stationary spots around the house where each spot will differ in terms of signal characteristics and the data from each testing spot will be combined to construct a scatter plot. The experiment will also be carried out during off-peak hours (0:00 to 3:00) in order to avoid performance inconsistencies due to load or interference as much as possible. The performance of four different MIMO configuration scenarios will be evaluated: 4×4, 4×2, 2×4 and 2×2 MIMO where the left-hand side digit corresponds to the number of active Tx ports at the eNB/gNB and the right-hand side digit corresponds to the number of active Rx ports at the UE. The dtac Band 28 cell serving the test location operates in 4T4R mode with DSS (Dynamic Spectrum Sharing) configured, allowing both LTE and NR to be deployed simultaneously on 10 MHz of bandwidth. Therefore, the 4×4 and 4×2 MIMO experiments will be carried out on the dtac network. The serving Advanced Info Service (AIS) Band 28 cell operates in 2T2R mode therefore the 2x4 and 2x2 MIMO experiments will be carried out on the AIS network. The bandwidth of the LTE and NR Band 28 carriers on AIS is 10 MHz and 15 MHz respectively, where 10 MHz is shared between LTE and NR but the remaining is reserved for NR.

For the maximum throughput test, the experiment will be carried out on the platform area in the MRT metro station in Bangkok. Indoor coverage inside the metro station is provided by the Distributed Antenna System (DAS). Each DAS node has two ports and are placed in pairs, allowing up to 4T4R operation as shown in Figure \ref{fig:DAS}. The DAS is shared by all three major operators and dtac is the only operator to deploy low-band 4×4 MIMO on Band 28. Since both Band 1 and Band 28 share the same DAS nodes, their 4×4 MIMO performance can be compared directly.

\begin{figure}[t!]
\centering
\begin{subfigure}{.24\textwidth}
  \centering
  \includegraphics[width=0.95\linewidth]{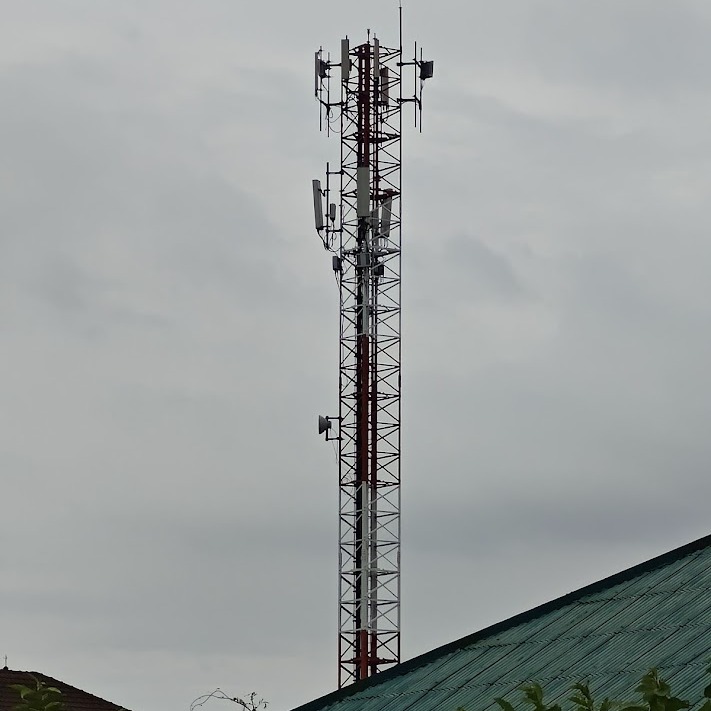}
  \vspace{-1mm}
  \caption{dtac}
  \label{fig:dtacTower}
\end{subfigure}%
\begin{subfigure}{.24\textwidth}
  \centering
  \includegraphics[width=0.95\linewidth]{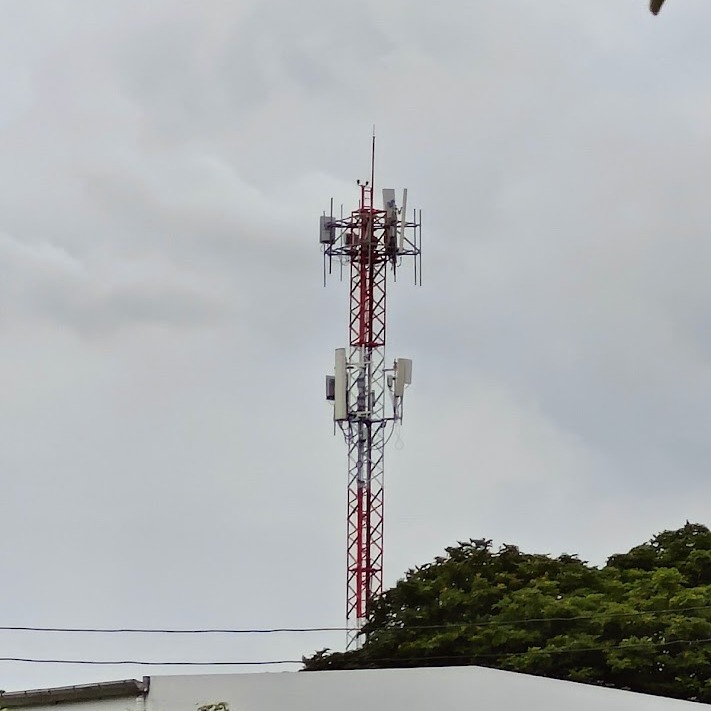}
  \vspace{-1mm}
  \caption{AIS}
  \label{fig:AISTower}
\end{subfigure}
\vspace{-2mm}
\caption{Serving cell tower for the reliability experiment}
\vspace{-4mm}
\label{fig:TowerPic}

\end{figure}

\begin{figure}
    \centering
    \includegraphics[width=0.87\linewidth]{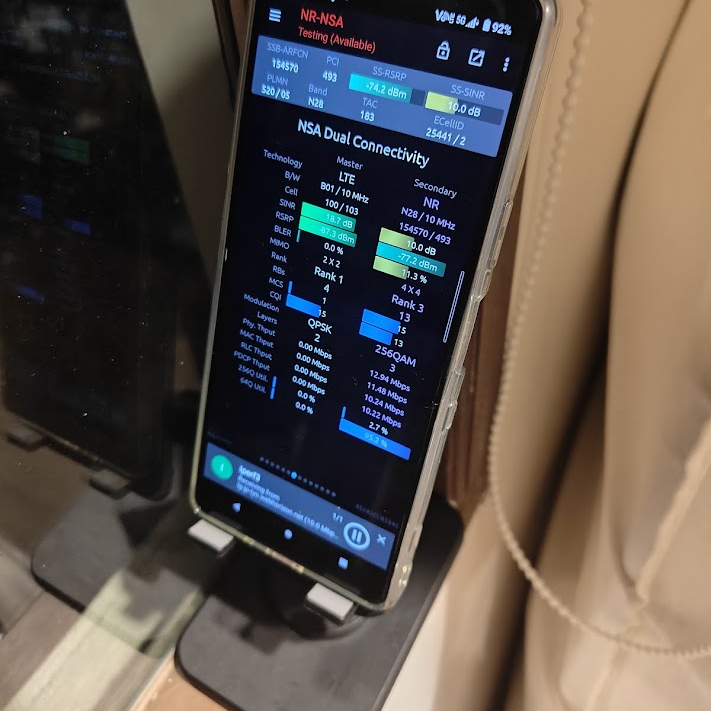}
    \caption{Test Setup}
    \label{fig:testSetup}
    \vspace{-7mm}
\end{figure}



\subsection{User equipment}
In this experiment, four different UEs are utilized which include the Sony Xperia 1 IV (XQT-72), Sony Xperia 1 III (SO-51B), ASUS Smartphone for Snapdragon insiders (EXP21), and OnePlus 11 (CPH2449). The Xperia 1 IV and the Xperia 1 III both support 4 Rx ports for Band 1, Band 3, and Band 28 simultaneously. For the other 2 UEs, 4 Rx ports are utilized for Band 1 and Band 3 but Band 28 is limited to 2 Rx ports. The performance of the Xperia 1 IV operating with 2 Rx ports will also be evaluated to obtain their relative 2 Rx mode performance to compare it with the OnePlus 11 and the ASUS EXP21 smartphone. To do this, 2 Rx ports will be disabled by applying firmware modification to the modem. Unfortunately, the modification cannot be done on the Xperia 1 III smartphone due to software restrictions.

\subsection{Data collection and analysis}
The performance is evaluated by collecting and analyzing the data that are reported by the modem. The modem data is collected using Network Signal Guru (NSG), a mobile network drive test application, which captures and decodes modem-reported data into a replayable log file. The NSG application records the reported modem parameters every 0.5-second intervals. The log file is then opened in the AirScreen software, where relevant parameters such as the Reference Signal Received Power (RSRP), Signal to Interference Noise Ratio (SINR), and Modulation and Coding Scheme (MCS) are selected and exported with corresponding timestamps to a CSV file. The CSV file is then opened in Microsoft Excel for data analysis and figure construction. \looseness=-1

The experiment will examine two use cases: the reliability test will assess the real-world performance of each device and the maximum performance test, where the Sony Xperia 1 IV device will be tested in a relatively optimal RF environment with excellent isolation to evaluate the maximum performance of LTE B28 4×4 MIMO compared to LTE B1 4×4 MIMO.

For the reliability test carried out at home, an Iperf3 loading script is utilized, which will run for 10 seconds at each test spot, applying a constant load. The iPerf3 server is hosted on a computer with port forwarding enabled. For the experiment carried out on the dtac network, the constant load is configured to 20 Mbps, and on the AIS network, it is set to 10 Mbps. The decision for these values is based on the guaranteed throughput that can be safely achieved across all testing spots within the house. \looseness=-1

For the maximum throughput test, the experiment will be carried out on the platform area of the Samyan, Wat Mangkorn, and Sanam Chai MRT metro stations in Bangkok.  These stations were chosen for their relatively low crowd density, which helps to minimize performance fluctuations due to congestion. In this experiment, the data is collected by doing a 2-minute walk test around the platform area while running a best-effort test script so that the maximum throughput can be evaluated. The test script used in this experiment is an HTTP GET script that establishes a connection to an Ookla speed test server hosted by AIS with a maximum bandwidth of 100 Gbps. The frequency bands that will be investigated are Band 1 and Band 28 and the performance will be evaluated by analyzing the PDCP throughput, SINR, MCS, RI, and 256QAM Utilization. 

\begin{figure}
    \centering
    \includegraphics[width=0.85\linewidth]{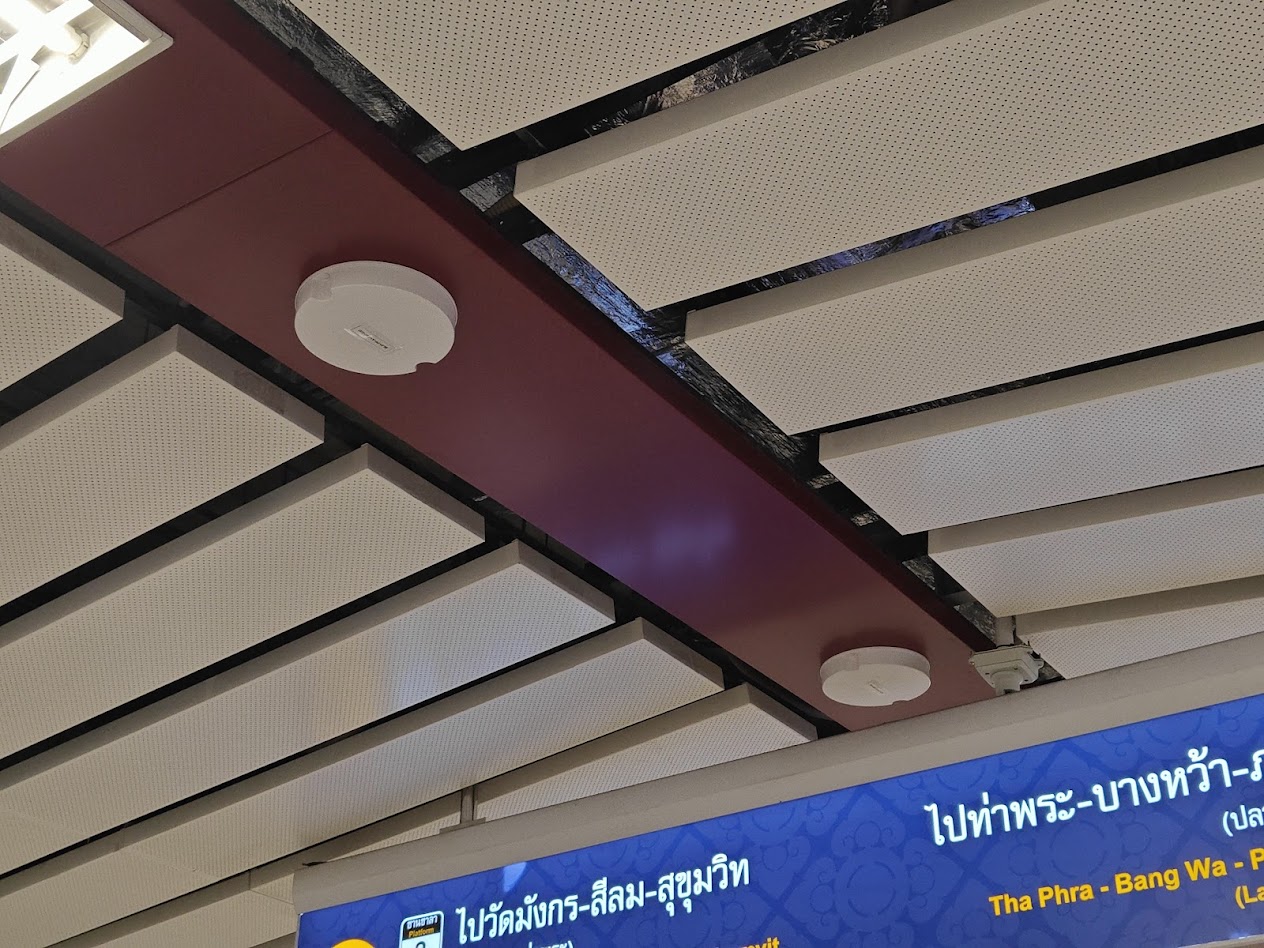}
    \caption{DAS nodes deployed in pairs for up to 4T4R operation}
    \label{fig:DAS}
    \vspace{-6mm}
\end{figure}

\section{Results and analysis}
\subsection{Reliability Test on NR Band 28 NSA}
\begin{figure}[t!]
\centering
\begin{subfigure}{.48\textwidth}
  \centering
  \includegraphics[width=0.95\linewidth]{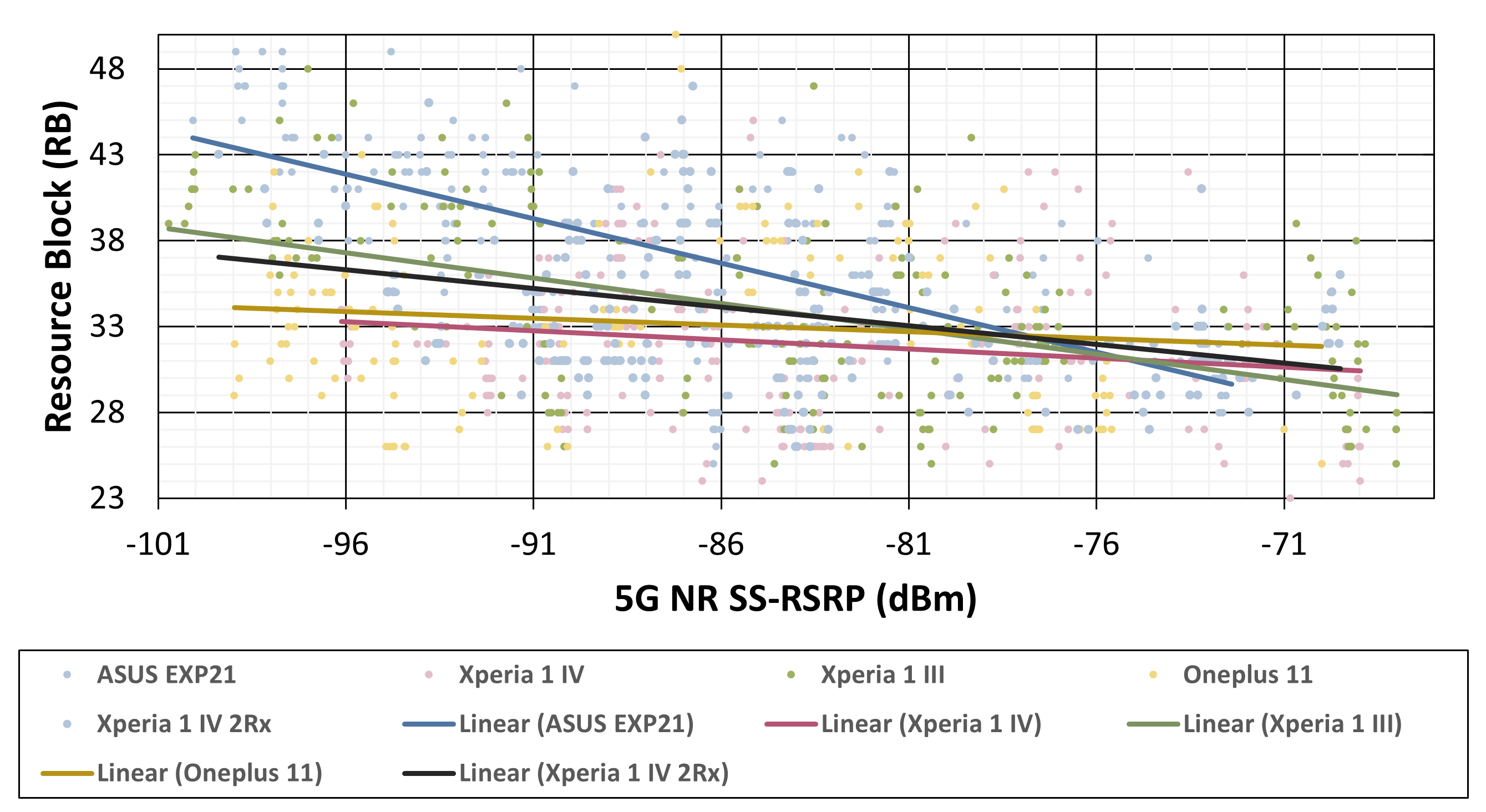}
  \vspace{-1mm}
  \caption{dtac}
  \label{fig:dtacN28Scatter}
\end{subfigure}\\
\begin{subfigure}{.48\textwidth}
  \centering
  \includegraphics[width=0.95\linewidth]{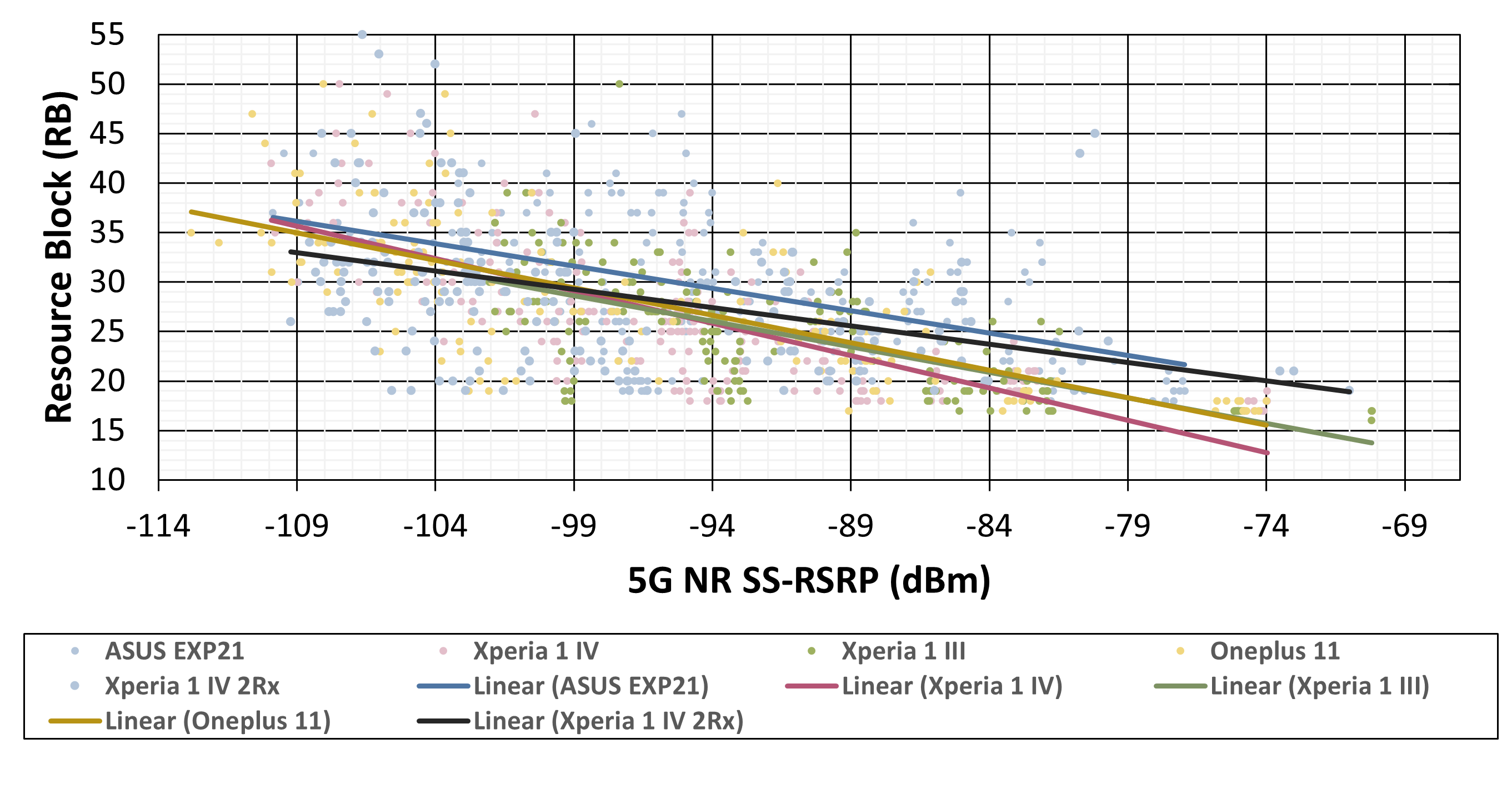}
  \vspace{-2mm}
  \caption{AIS}
  \label{fig:AISN28Scatter}
\end{subfigure}
\caption{Scatter Plot of Resource Block (RB) vs NR SS-RSRP}
\vspace{-4mm}
\label{fig:NRScatter}
\vspace{-3mm}
\end{figure}

Starting with the results of the 4×4 and 4×2 MIMO scenario on the dtac network, the gains with low-band 4 Rx in a 4×4 MIMO scenario showed the largest performance difference, especially under low RSRP conditions where the biggest difference across all four devices was observed as shown in Figure \ref{fig:dtacN28Scatter} where there is a clear divergence of the regression lines. However, under better signal conditions, the performance difference begins to diminish and eventually becomes negligible. The results show that the implementation of 4 Rx on low-band is effective at improving user experience under poor signal conditions, just like the 4 Rx implementation on mid-band frequencies \cite{huawei20204x4mimo}. While the results suggest that there are little to negligible performance gains under good signal conditions, when under weak signal conditions, especially where the RSRP falls below -100dBm, the Sony Xperia 1 IV with 4 Rx is able to negotiate a higher MCS on average, thus requiring less RB allocation in order to maintain the 20 Mbps throughput target. Furthermore, when the Sony Xperia 1 IV device was forced to utilize 2 Rx ports, a clear performance decline was observed. The performance of the OnePlus 11 is very comparable to the Xperia 1 IV smartphone, despite utilizing only 2 Rx ports. On the Sony Xperia 1 III smartphone, the performance appears to be comparable to the Xperia 1 IV smartphone operating with 2 Rx ports and slightly behind the OnePlus 11, which shows that 4 Rx port configuration does not always outperform 2 Rx port configurations as the design and characteristics of the antenna still have a big influence on the overall performance \cite{Bancroft2004FundamentalDL}.

\begin{figure}[t!]
\centering
\begin{subfigure}{.48\textwidth}
  \centering
  \includegraphics[width=0.95\linewidth]{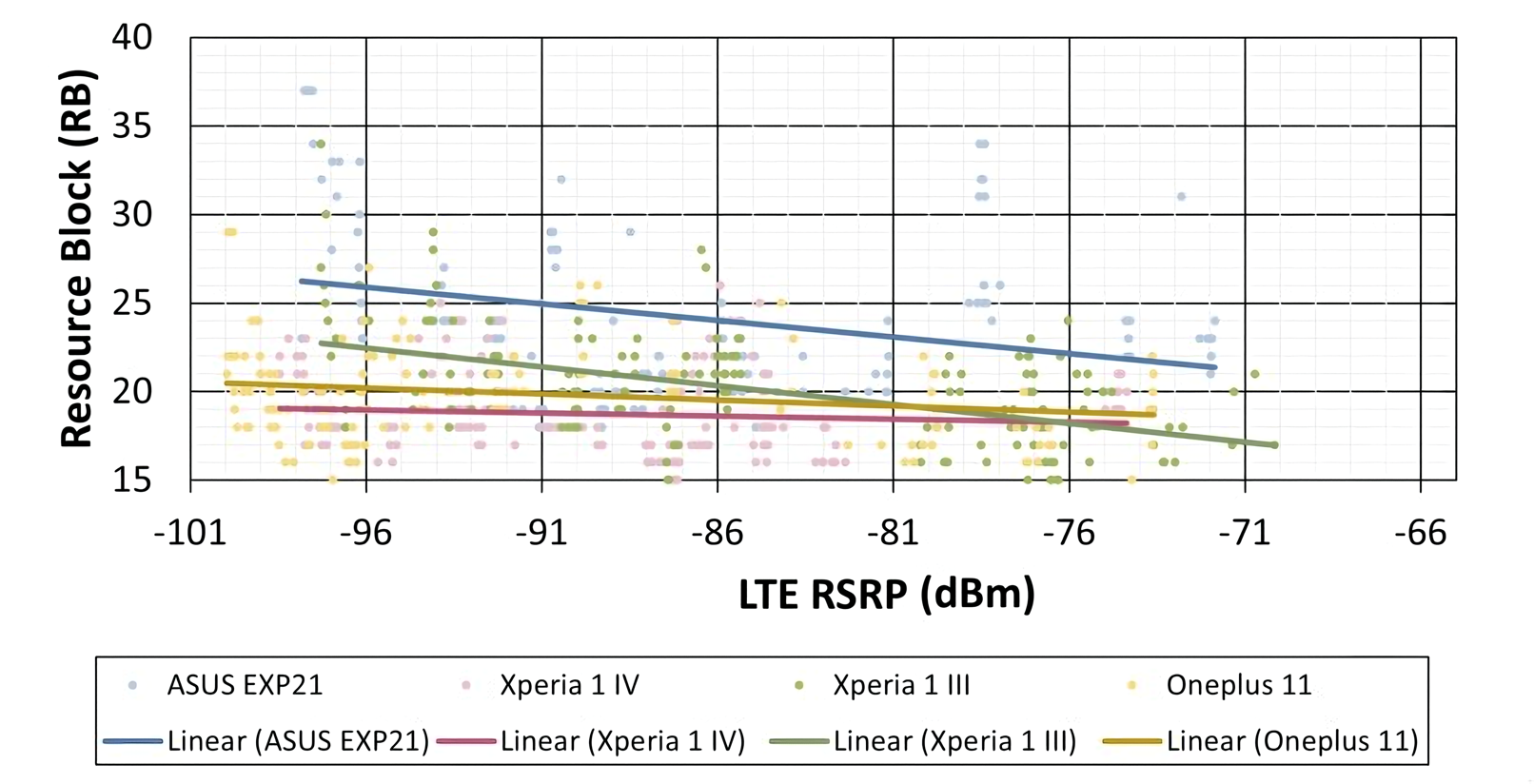}
  \vspace{-1mm}
  \caption{dtac}
  \label{fig:dtacB28Scatter}
\end{subfigure}\\
\begin{subfigure}{.48\textwidth}
  \centering
  \includegraphics[width=0.95\linewidth]{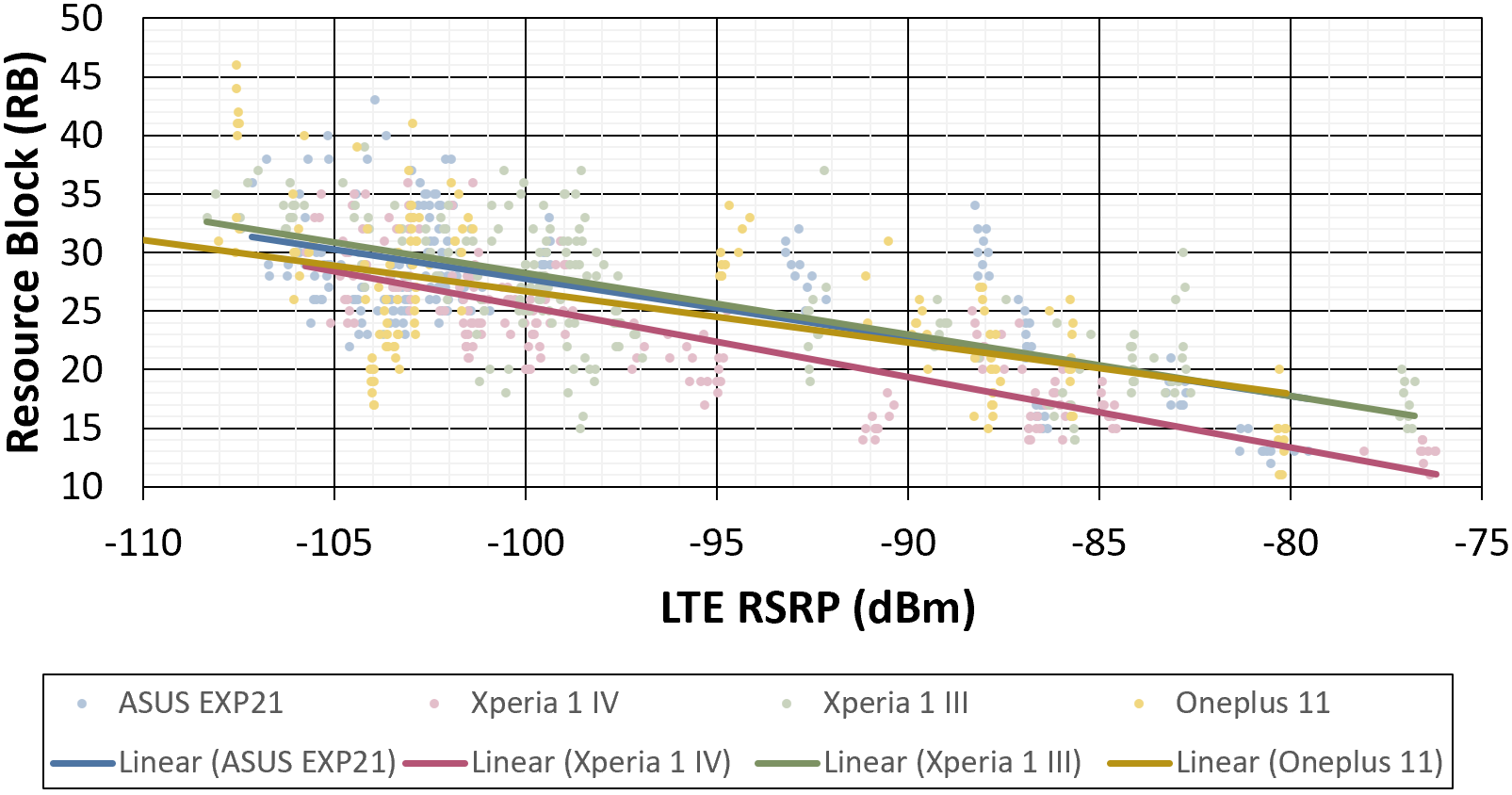}
  \vspace{-1mm}
  \caption{AIS}
  \label{fig:AISB28Scatter}
\end{subfigure}
\caption{Scatter Plot of Resource Block (RB) vs LTE RSRP}

\label{fig:LTEScatter}

\end{figure}

Moving on to the 2×4 and 2×2 MIMO scenario, the results showed less noticeable performance gains for both Xperia 1 IV and Xperia 1 III devices at all signal conditions where the regression line for all of the devices have a similar gradient that almost overlap each other as shown in Figure \ref{fig:AISB28Scatter}. Unlike in the results of the 4×4 MIMO scenario, a clear performance gain at low RSRP conditions is not observed, showing that the maximum benefit of 4 Rx antenna in phones can only be realized when 4 Tx ports are utilized at the eNB/gNB. Similar to the 4×4 MIMO scenario, the Xperia 1 IV showed generally lower RB utilization at all RSRP levels. Lastly, just like in the experiment on the dtac network, when the Xperia 1 IV was forced to operate with 2 Rx ports, a measurable decline in performance was observed as well.


\subsection{Reliability Test on LTE Band 28}

The results of the experiment on LTE showed similar results that closely parallel with the experimental results of NR. Figure 8 shows the results of the 4×4 and 4x2 MIMO scenarios on the dtac LTE Band 28. Similar to the results on NR, the Sony Xperia 1 IV device is able to maintain a relatively stable RB utilization across the RSRP range and is the top performer overall. Just like in the NR experiment, the performance of the OnePlus 11 smartphone is very similar to the Sony Xperia 1 IV smartphone, only underperforming slightly. The ASUS EXP21 smartphone showed the worst performance, just like in the NR experiment. The Xperia 1 III smartphone also showed results that parallel with the NR experimental where the device showed leading performance under strong signal conditions but experiences a more dramatic performance decline with decreasing RSRP, leading to middling performance under weak signal conditions.
Lastly, for the 2×4 and 2×2 MIMO scenario on the AIS LTE Band 28 network, the results also closely parallel with the NR results where the RB utilization of the Xperia 1 IV is generally slightly lower than other testing devices across the RSRP range, especially in stronger signal conditions but the performance improvement diminishes with weaker signals. For the other devices, the RB utilization is nearly indistinguishable from each other as shown in Figure 8 where the lines of regression closely overlap each other. When the Xperia 1 IV device was forced to operate with 2 Rx ports, there was also a clear decline in performance where it performed very similarly to the ASUS EXP21 smartphone. 


\subsection{MCG Limitation in Non-Standalone access}

\begin{figure*}[t!]
\centering
\begin{subfigure}{.32\textwidth}
  \centering
  \includegraphics[width=1\linewidth]{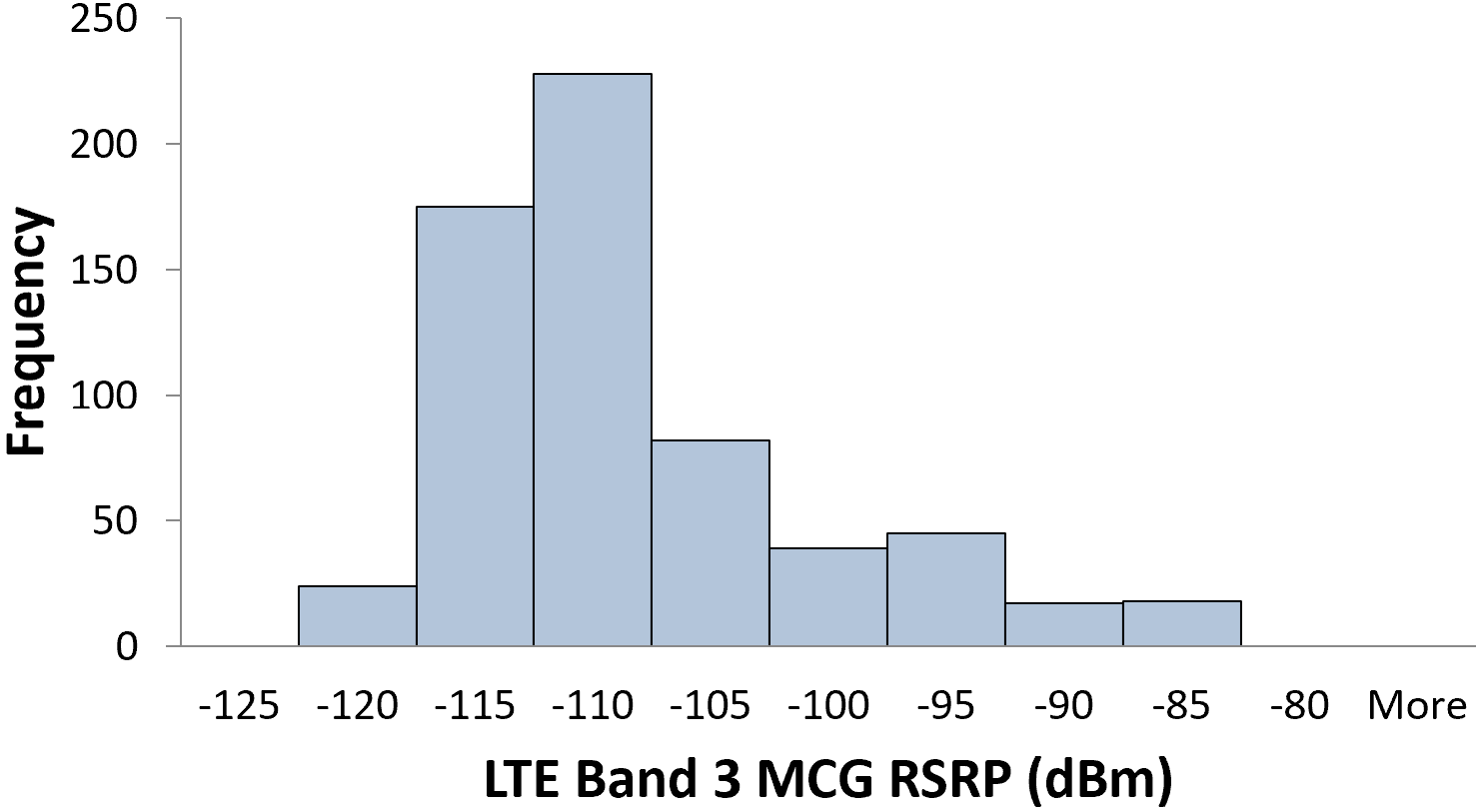}
  \caption{ASUS EXP21}
  \label{fig:EXP21Dist}
\end{subfigure}%
\begin{subfigure}{.32\textwidth}
  \centering
  \includegraphics[width=1\linewidth]{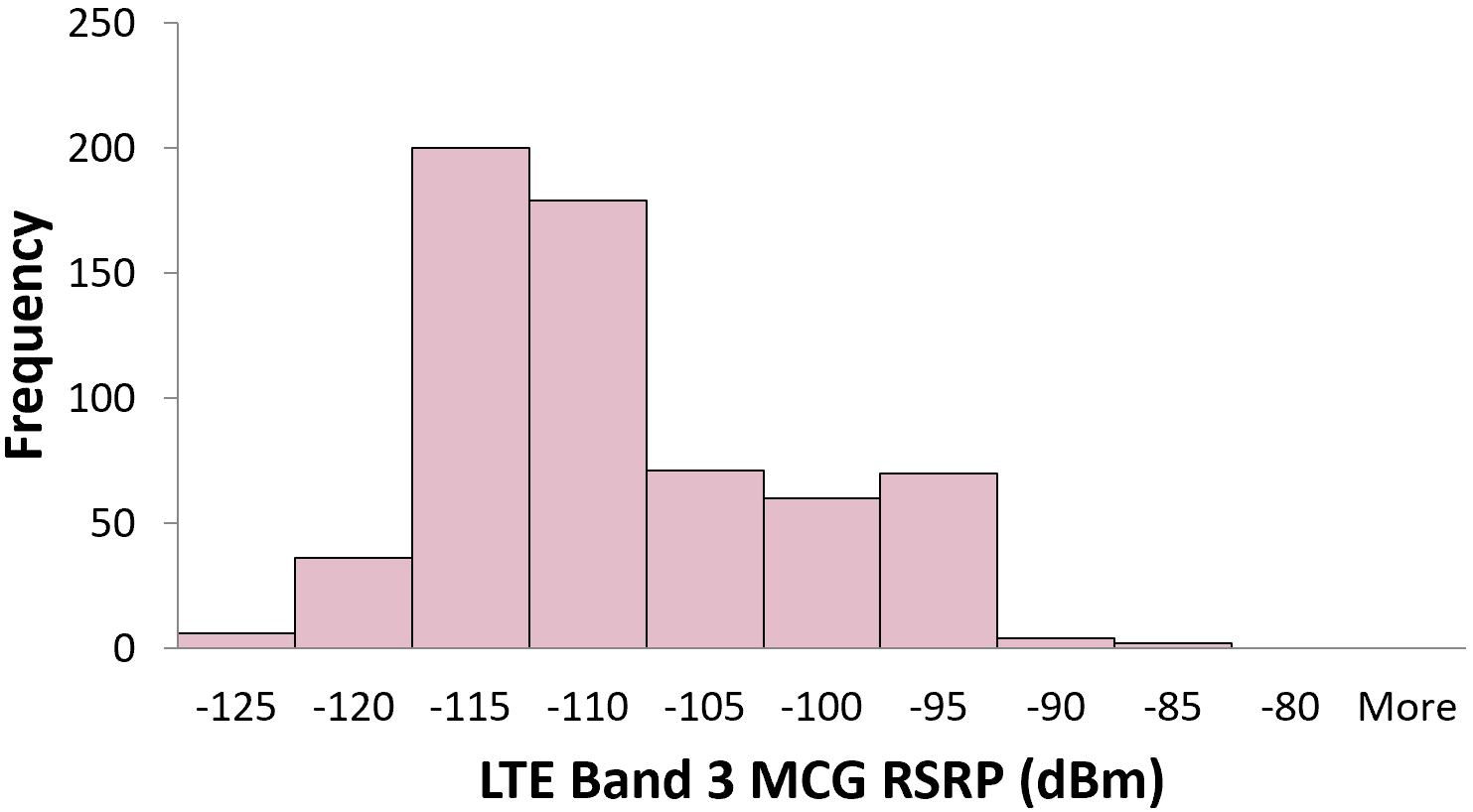}
  \caption{Xperia 1 IV}
  \label{fig:XperiaIVDist}
\end{subfigure}%
\begin{subfigure}{.32\textwidth}
  \centering
  \includegraphics[width=1\linewidth]{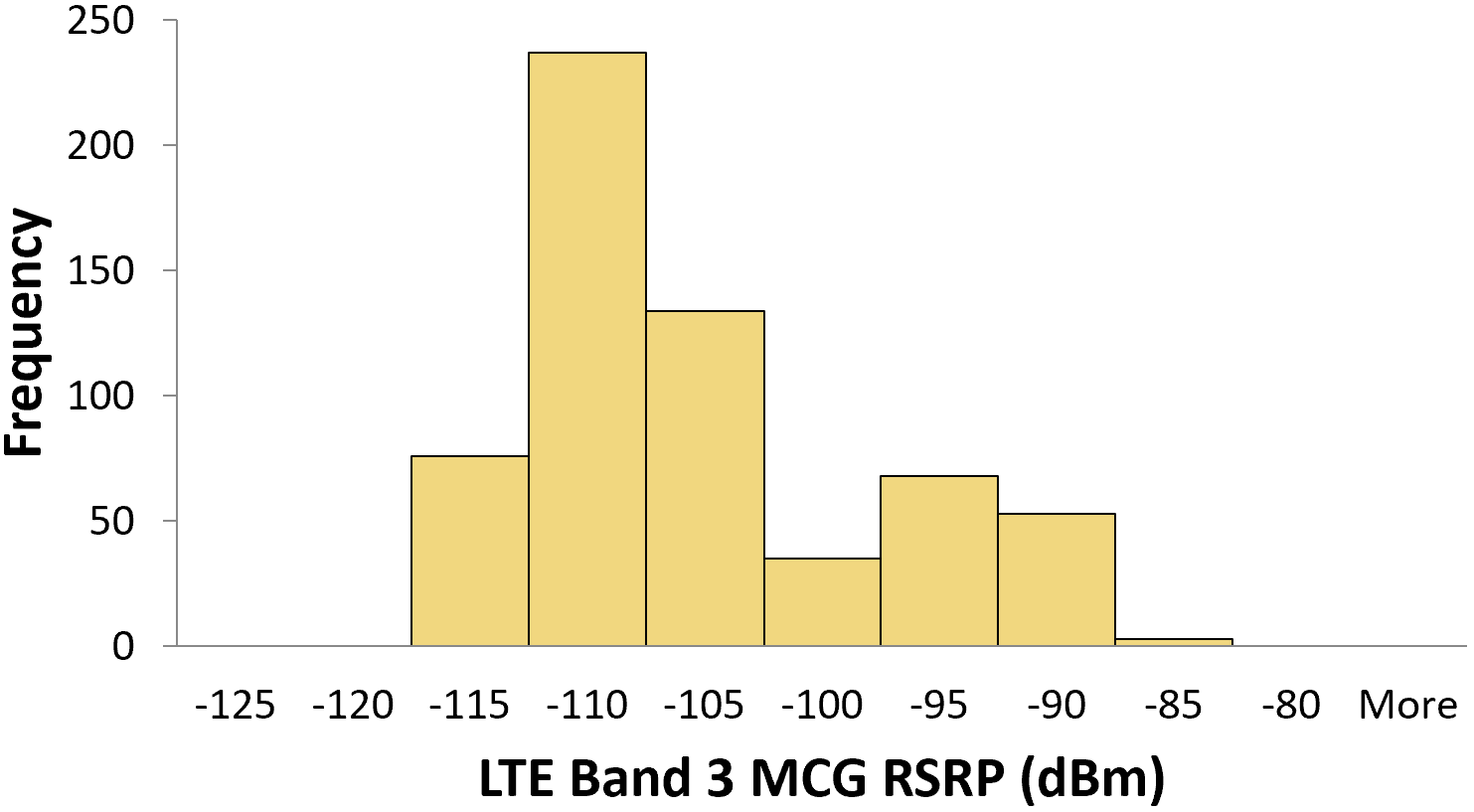}
  \caption{OnePlus 11}
  \label{fig:OnePlusDist}
\end{subfigure}

\vspace{-1mm}
\caption{LTE B3 RSRP distribution in the reliability test}
\vspace{-6mm}

\end{figure*}

As shown in the previous section, UEs equipped with 4 Rx on low-band can enhance user experience, especially under cell edge conditions and even more so when the eNB/gNB operates with 4 Tx ports. However, in Non-Standalone (NSA) option 3 access, the NR carrier acts as the Secondary Cell Group (SCG) and relies on the LTE Main Cell Group (MCG) for the control plane signaling. \cite{3GPP_23-501}. This is a problem for low-band NR because the LTE MCG will most likely be a band with higher frequency which experiences more attenuation. Therefore, ultimately bottlenecks the accessibility of low-band NR. This is not the case for NR standalone access because the low-band NR gNB operates independently. Figure \ref{fig:EXP21Dist}, \ref{fig:XperiaIVDist} and \ref{fig:OnePlusDist} show the cumulative distribution of LTE B3 MCG RSRP data points for ASUS EXP21, Xperia 1 IV and OnePlus 11, respectively, during the reliability experiment on the AIS network. The results show that the Xperia 1 IV smartphone exhibits worse LTE B3 reception as shown by the higher accumulation of LTE B3 RSRP at -115 dBm and below. At one point during the experiment, the LTE B3 RSRP on the Xperia 1 IV device fell below the -128 dBm threshold, triggering a RAT-fallback to 3G, an event not observed with the other devices. This outcome suggests that strong low-band reception does not necessarily correlate to better reception in other frequency bands. It's also important to note that the performance of the NR SCG carrier is independent of the LTE MCG carrier's reception quality, as long as the connection is maintained.

%

\subsection{Maximum throughput test}

\begin{table}[!t]
\setstretch{0.9}
\vspace{-3mm}
\caption{Indoor Walk Test Band 1 (2100 MHz)}
\vspace{-1mm}
\label{tab:performance_metrics_B1}
\centering
\resizebox{7cm}{!}{\begin{tabular}{cccc}
\toprule
 & & Train Station & \\
& BL27 & BL29 & BL31 \\
\midrule
Mean PDCP Thput (Mbps) & 122.33 & 140.93 & 151.86 \\
Mean SINR (dB) & 32.57 & 27.06 & 32.48 \\
Mean MCS & 22.96 & 25.58 & 25.67 \\
256QAM Utilization (\%) & 79.08 & 99.36 & 98.14 \\
Rank 4 Utilization (\%) & 75.09 & 62.43 & 75.57 \\
Rank 3 Utilization (\%) & 17.70 & 19.99 & 19.77 \\

\bottomrule
\end{tabular}}
\vspace{-3mm}
\end{table}

\begin{table}[!t]
\setstretch{0.9}
\caption{Indoor Walk Test Band 28 (700 MHz)}
\vspace{-1mm}
\label{tab:performance_metrics_B28}
\centering
\resizebox{7cm}{!}{\begin{tabular}{cccc}
\toprule
 & & Train Station & \\
 & BL27 & BL29 & BL31 \\
\midrule
Mean PDCP Thput (Mbps) & 112.43 & 114.90 & 116.69 \\
Mean SINR (dB) & 32.47 & 34.71 & 32.67 \\
Mean MCS & 26.02 & 25.90 & 25.92 \\
256QAM Utilization (\%) & 98.79 & 99.42 & 99.05 \\
Rank 4 Utilization (\%) & 2.40 & 0 & 6.34 \\
Rank 3 Utilization (\%) & 93.32 & 88.41 & 85.65 \\
\bottomrule
\end{tabular}}
\vspace{-6mm}
\end{table}

The indoor 120-second walk test with unlimited bandwidth shows that LTE Band 1 has a higher average throughput due to much higher utilization of Rank 4 MIMO compared to LTE Band 28 with an average rank 4 utilization of 71.03\% and 2.91\% respectively as shown in Table \ref{tab:performance_metrics_B1} and \ref{tab:performance_metrics_B28}. On the other hand, Rank 3 was mostly utilized on Band 28. The results also show that the mean MCS of B1 is slightly lower than B28 in all stations. This phenomenon could be explained by the higher utilization of Rank 4. The overall average throughput of B1 and B28 is 138.37 Mbps and 114.67 Mbps respectively, which still exceeds the theoretical throughput of Rank 2 MIMO which is around 100 Mbps. The results show that the Xperia 1 IV device struggles to negotiate a higher order MIMO rank on low-band despite the identical DAS deployment of B1 and B28. However, under more ideal conditions, it is possible to maintain a high Rank 4 utilization and achieve similar results as Band 1 (see Table \ref{tab:performance_metrics_B1}). This was verified after running the same test script in an optimal spot in Phetchaburi station (BL21) while standing directly below one of the DAS nodes while keeping the device stationary. The test script was run for 75 seconds and The average Rank 4 utilization was 55.29\% with an average throughput of 140.35 Mbps. Overall, the results suggest that the throughput and capacity gains on low-band with 4×4 MIMO will be more limited compared to higher frequency bands such as Band 1 due to the fact that it is more difficult to sustain higher MIMO orders.

\section{Conclusion}
The experiment shows that low-band 4×4 MIMO in smartphones provides the same benefits observed with 4×4 MIMO on higher frequency bands such as enhancing reliability under cell-edge conditions and enabling higher throughput that exceeds the 2-Layer MIMO theoretical maximum with Rank 3 and Rank 4 utilization under good signal conditions. However, the result suggests that the 4×4 MIMO performance gains scale better with higher frequencies. When the Xperia 1 IV device was forced to utilize 2 Rx ports, a clear decline in performance was observed in the real-world reliability test, confirming that two additional Rx ports provide real performance benefits. However, the results also demonstrate that UEs equipped with 4 Rx on low-band do not always outperform UEs with 2 Rx ports configuration, as shown by the results of the Xperia 1 III device where the OnePlus 11 smartphone consistently outperforms it in the real-world reliability test for both LTE and NR despite utilizing 2 Rx ports. The low-band performance of the OnePlus 11 smartphone was found to be nearly as good as the Xperia 1 IV with 4 Rx ports, demonstrating that it is possible to achieve Xperia 1 IV-like performance without the need for two additional Rx ports in real-world scenarios. However, 4 Rx ports are still required for higher theoretical throughput. When it comes to maximum throughput and capacity gains, the results shows that it is harder to negotiate higher order MIMO rank on low-band compared to higher frequencies bands such as Band 1, which suggests that the implementation of 4×4 MIMO in low-band will offer limited gains compared to higher frequency bands for these applications. Ultimately, the goal of this research is to provide insights into the real-world impact of low-band 4×4 MIMO implementation in commercial UEs, which could guide the UE manufacturers in terms of decision-making related to antenna design. Moreover, the data obtained from this experiment can serve as a valuable benchmark for mobile network operators, aiding their decision-making processes for infrastructure upgrades and future network rollout.

\section*{Acknowledgement}
This paper is supported by the commissioned research JPJ012368C03801, National Institute of Information and Communications Technology (NICT), Japan. Additionally, the authors would like to express their gratitude to PEI Xiaohong of Qtrun Technologies for providing the Network Signal Guru (NSG) and AirScreen software, which was utilized to collect data reported by the modem in real-time and replaying them afterward for analysis, which greatly facilitated the completion of this research. The first author would also like to thank the Department of Computer Science and Communications Engineering, Waseda University for providing the Sony Xperia 1 III (SO-51) smartphone and the ASUS Snapdragon Insiders Smartphone (EXP21). Finally, special thanks are extended to \textbf{Associate Professor Chaodit Aswakul, Ph.D} at Chulalongkorn University for guidance and support throughout the research. \looseness=-1

\setstretch{0.95}
\Urlmuskip=0mu plus 1mu\relax
\bibliographystyle{IEEEtran}
\bibliography{References}

@INPROCEEDINGS{10118777,
  author={Arunruangsirilert, Kasidis and Wongprasert, Pasapong and Katto, Jiro},
  booktitle={2023 IEEE Wireless Communications and Networking Conference (WCNC)}, 
  title={Performance Evaluations of C-Band 5G NR FR1 (Sub-6 GHz) Uplink MIMO on Urban Train}, 
  year={2023},
  volume={},
  number={},
  pages={1-6},
  doi={10.1109/WCNC55385.2023.10118777}}

@standard{3GPP_36-306,
type = {Standard},
title = {Evolved Universal Terrestrial Radio Access (E-UTRA); LTE physical layer; General description},
language = {en},
number = {TS 36.306 Version 16.0.0 Release 16},
institution = {3GPP},
year = {Jul. 2020}}

@online{AppleiPhoneSESpecs,
    author = {{Apple Inc.}},
    title = {iPhone SE - Technical Specifications},
    organization = {Apple Inc.},
    url = {https://www.apple.com/iphone-se/specs/},
    year = {2023},
    note = {Accessed on 24 October 2023}
}

@inproceedings{Bancroft2004FundamentalDL,
  title={Fundamental Dimension Limits of Antennas Ensuring Proper Antenna Dimensions in Mobile Device Designs},
  author={Randy Bancroft},
  year={2004},
  url={https://api.semanticscholar.org/CorpusID:17860392},
  note={Accessed on 24 October 2023}
}

@standard{3GPP_38-212,
    type = {Standard},
    title = {NR; Multiplexing and channel coding},
    language = {en},
    number = {TS 38.212 Version 17.6.0 Release 17},
    institution = {3GPP},
    year = {Jun. 2023}
}

@standard{3GPP_36-212,
    type = {Standard},
    title = {Evolved Universal Terrestrial Radio Access (E-UTRA); Multiplexing and channel coding},
    language = {en},
    number = {TS 36.212 version 15.2.1 Release 15},
    institution = {3GPP},
    year = {Jun. 2023}
}

@standard{3GPP_36-213,
    type = {Standard},
    title = {Evolved Universal Terrestrial Radio Access (E-UTRA); Physical layer procedures},
    language = {en},
    number = {TS 36.213 version 10.4.0 Release 10},
    institution = {3GPP},
    year = {Jun. 2012}
}

@standard{3GPP_23-501,
    type = {Standard},
    title = {5G; System architecture for the 5G System (5GS)},
    language = {en},
    number = {TS 23.501 version 10.4.0 Release 16},
    institution = {3GPP},
    year = {Jun. 2020}
}

@online{huawei20204x4mimo,
  author = {Huawei},
  title = {4x4 mimo: The key to unlocking 5g potential},
  year = {2020},
  url = {https://www-file.huawei.com/-/media/corporate/pdf/mbb/2020/4x4-mimo.pdf?la=en},
  note = {Accessed on 26 October 2023}
}

\end{document}